\documentclass[manuscript]{acmart}

\AtBeginDocument{%
  \providecommand\BibTeX{{%
    \normalfont B\kern-0.5em{\scshape i\kern-0.25em b}\kern-0.8em\TeX}}}

\renewcommand\footnotetextcopyrightpermission[1]{}
\copyrightyear{}
\acmYear{}
\acmDOI{}
\acmConference[]{}{}{}
\acmBooktitle{}
\acmPrice{}
\acmISBN{}

\begin{document}

\title[Diversity in the Music Listening Experience]{Diversity in the Music Listening Experience: Insights from Focus Group Interviews}

\author{Lorenzo Porcaro}
\email{lorenzo.porcaro@upf.edu}
\affiliation{%
  \institution{Music Technology Group, Universitat Pompeu Fabra}
  \city{Barcelona}
  \country{Spain}
}

\author{Emilia G\'{o}mez}
\email{emilia.gomez@upf.edu}
\affiliation{%
  \institution{Music Technology Group, Universitat Pompeu Fabra}
  \city{Barcelona}
  \country{Spain}
}
\affiliation{%
  \institution{Joint Research Centre, European Commission}
  \city{Sevilla}
  \country{Spain}
}

\author{Carlos Castillo}
\email{carlos.castillo@upf.edu}
\affiliation{%
  \institution{Web Science and Social Computing Research Group, Universitat Pompeu Fabra}
  \city{Barcelona}
  \country{Spain}
}
\affiliation{%
  \institution{ICREA}
  \city{Barcelona}
  \country{Spain}
}
\renewcommand{\shortauthors}{Porcaro, et al.}

\begin{abstract}
Music listening in today's digital spaces is highly characterized by the availability of huge music catalogues, accessible by people all over the world. 
In this scenario, recommender systems are designed to guide listeners in finding tracks and artists that best fit their requests, having therefore the power to influence the diversity of the music they listen to.
Albeit several works have proposed new techniques for developing diversity-aware recommendations, little is known about how people perceive diversity while interacting with music recommendations. 
In this study, we interview several listeners about the role that diversity plays in their listening experience, trying to get a better understanding of how they interact with music recommendations. 
We recruit the listeners among the participants of a previous quantitative study, where they were confronted with the notion of diversity when asked to identify, from a series of electronic music lists, the most diverse ones according to their beliefs.
As a follow-up, in this qualitative study we carry out semi-structured interviews to understand how listeners may assess the diversity of a music list and to investigate their experiences with music recommendation diversity. 
We report here our main findings on 1) what can influence the diversity assessment of tracks and artists' music lists, and 2) which factors can characterize listeners' interaction with music recommendation diversity. 
\end{abstract}




\maketitle

\section{Introduction}\label{sec:introduction}
Music Recommender System (Music RS) research is one of the Music Information Retrieval (MIR) areas wherein scientific contributions have been translated into a widespread technology, especially in commercial settings, and with potential implications for millions of listeners all around the world \cite{Jannach2020, Salamon2019}. 
Despite their large adoption in streaming services, little is known about how listeners, the Music RS end-users, perceive the interaction with music recommendations.
Among the aspects characterizing these interactions, the diversity of the music that listeners are exposed to has been shown to affect the perceived quality of the recommendations \cite{Knees2019}.

Several definitions of diversity can be found in the scientific and humanistic literature, which according to the context can be measured differently and with different ethical-epistemic implications, as discussed by Steel et al. in \cite{Steel2018}.
Analyzing the relationship between listeners and Music RS, diversity may be traced back on the one side to the features of the recommended items (e.g. tracks, playlists, or artists), on the other side to the characteristics of the users exposed to such items, and moreover to the interaction between users and items \cite{Porcaro2021}.
Whilst the latter aspect has been the focus of most of the computational approaches proposed in the Music RS literature, in this work we center our attention on the concept of \textit{perceived diversity}, intended as the diversity of the items as perceived by the users.
Under this lens, one of the goals of this study is to deepen what are the characteristics that listeners consider relevant while evaluating the diversity of a music list formed by tracks or artists, focusing around two principal concept clusters, of musical variation and socio-demographic variation.

In order to do this, we interview several listeners on the role that diversity plays in their everyday listening experience. 
We recruit them among the participants of a previous user study \cite{Porcaro2022}, in which they were confronted with the concept of diversity when asked to identify which among a series of lists, formed by Electronic Music (EM) tracks and artists, they perceived to be more diverse. 
The goal of this study is to complement such findings with a qualitative analysis by means of focus group interviews. 
As a result of conducting a qualitative content analysis \cite{Philipp2000} on the experiences shared by the interviewees, we argue how a lack of prior knowledge of a specific music culture may increase the difficulty of the diversity assessments. 
Indeed, people unfamiliar with a kind of music would mostly rely on acoustic features (e.g. rhythm) or stereotypical representations of such music (e.g. EM tracks have fast tempo) for evaluating the diversity of the lists. 
On the contrary, to be familiar with a music genre may lead to a faster categorization of tracks and artists, facilitating the assessment but at the same time inducing bias deriving from prior knowledge. 
We complement these findings by discussing the role that music recommendation diversity may have in the listeners' experience, reinforcing the idea that diversity needs can vary among individuals, also highlighting the positive and negative factors that can influence the listeners. 

The rest of the paper is organized as follows. 
In Section \ref{sec:background}, we briefly review the MIR literature on diversity and music listening habits, focusing on user studies analyzing the concept of perceived diversity. 
Section \ref{sec:method} summarizes the considered methodology, whilst Section \ref{sec:insights} presents the main findings from the interviews. 
Finally, conclusions and future work are drawn in Section \ref{sec:conclusions}. 
\section{Background}\label{sec:background}
Scholars in fields such as Information Retrieval (IR) and Human-Computer Interaction (HCI) continue to question how diversity can be operationally defined \cite{Amigo2018}, how it is perceived by information systems' users \cite{Sakai2019}, and what are the implications to be considered when designing diversity metrics \cite{Mitchell2020}. 
Such foundational research is still missing in the MIR and Music RS literature, which instead has mostly focused on empirical analysis \cite{Porcaro2021}.

Nonetheless, several works in the MIR literature have already investigated the relation between diversity needs and personality \cite{Lu2018}, individual traits \cite{Jin2020}, country's cultural dimensions \cite{Ferwerda2016}, and socio-economic status \cite{Park2016}. 
For instance, Robinson et al. in \cite{Robinson2020} study users' perception of diversity, investigating its connection with personalization techniques.
The authors characterize the presence of inner and outer diversity, the former related to the music that users usually listen to, while the latter the one far from their preferences.
Liu et al. in \cite{Lu2018} explore the relationships between diversity needs and personality traits, showing how the diversification technique based on such traits can improve the overall satisfaction for the received music recommendation. 
Similarly, Jin et al. in \cite{Jin2020} consider how users’ interface to navigate diversity-aware music recommendations can influence differently the perceived diversity, highlighting the role that individual traits may have in such interactions. 
Perceived diversity is also investigated by Ferwerda et al. in \cite{Ferwerda2017}, where the authors show its positive influence on the discovery of music and recommendation attractiveness.  
These findings are confirmed by the exploratory study in a commercial system developed by Tintarev et al. in \cite{Tintarev2017}, where participants confirm that diversity, together with novelty, are two valuable factors while interacting with recommendations designed to favor music exploration. 
Furthermore, through a user study Kim et al. in \cite{Kim2020} compare the impact that music recommendations from interpersonal (e.g. friends) and non-interpersonal (e.g. systems) channels may have, highlighting how diversity has a significant effect on the adoption of recommendations from non-interpersonal channels.

Undoubtedly, these results can foster the discussion about the different perceptions of diversity of music recommendation, but the absence of a general framework for studying music recommendation diversity makes it difficult to interpret such results under a more comprehensive perspective \cite{Porcaro2019}. 
Following previous research, this study analyzes qualitatively how the concept of diversity and its perception can be interpreted by listeners, focusing on a specific musical genre, Electronic Music.
\section{Methodology}\label{sec:method}
The present work is a follow-up to a user study where subjects were asked to select which among two music lists they perceived to be more diverse. 
The study is summarized as follows.
First, they had to listen to two lists of EM audio tracks and select the most diverse in terms of music features. 
Afterwards, they had to compare two lists of four EM artists' photos and select the most diverse list according to their social salient attributes (e.g. gender, skin tone, age). 
Finally, we mixed the former tasks presenting two lists formed by four artists’ photos and corresponding tracks, and participants had to consider both tracks' musical features and artists' attributes when assessing the diversity of the lists.
Regarding the material included in the study, we manually searched tracks and artists' data in several websites such as \textit{MusicBrainz}, \textit{Wikipedia}, \textit{AllMusic}, and also in artists' websites and electronic music magazines. 
The anonymized responses, a facsimile of the survey designed for the study, and the complete list of tracks and artists presented are publicly available,\footnote{\url{https://zenodo.org/record/4436649}} and for a comprehensive description of the study we refer to \cite{Porcaro2022}.
The study was an activity of the TROMPA project,\footnote{An international, publicly-funded research project: \url{https://trompamusic.eu/}} and followed the personal data management and ethical protocols approved within this project.

As a follow-up, we conducted 7 semi-structured interviews with 14 participants of the study, divided into focus groups of 2-3 participants when possible (P2 and P3; P4 and P5; P6, P7, and P8; P9 and P10; P11, P12, and P13), if not interviewing them individually (P1; P14). 
We choose to collect qualitative data through focus group interviews because the interviewees shared the experience of having previously participated in the aforementioned study. 
This sense of belonging to a group has been proven to facilitate the participants in feeling safe when sharing their opinions, making it possible to collect spontaneous data occurring thanks to the interaction among them \cite{Onwuegbuzie2009}. 
Interviewees were balanced in terms of gender (8 male, 7 female), mostly from Europe (86\%), aged 18-45, and all of them hold a bachelors' degree or higher. 
The majority declared to have received formal music training (71\%), while only a few of them affirmed to regularly engage in playing, DJing or producing any kind of music (35\%). 
Regarding their listening habits, most of them described their musical taste as varied (79\%), while less declared to listen often to EM (71\%), and even less within EM described their musical taste as varied (57\%).

The interview was divided into two main parts, lasting for a total of one hour for the focus groups, and half an hour for the individual interviews.
In the first part, we asked participants to discuss their strategy to assess the diversity of a music list while participating in the study.
In the second part, we centered the discussion around participants' experiences with Music RS, and the role that diversity may play in that specific scenario. 
Interviews were conducted in English, Spanish and Italian, and then transcribed in English.
We started the analysis of the transcripts by doing open coding, annotating the several themes and categories touched by the interviewees. 
Following, a process of axial coding was performed to connect the distinct categories and identify the shared themes, finally collapsing our findings into two main themes, reported in the next section.
\section{Insights from Interviews}\label{sec:insights}

\subsection{Assessing the Diversity of a Music List}\label{sec:qual1}

Interviewees mentioned several aspects related to what diversity means to them, and how it has been interpreted while comparing lists of tracks and artists.
When asked about their strategy in choosing the most diverse lists, they often started the discussion by self-identifying themselves as being or not experts in EM. 

Those who do not identify as an expert of this genre highlighted how the diversity assessment has been mostly based on listenable characteristics. 
Some of them also recognized how difficult it was to assess the diversity of the lists:
``\textit{It was very difficult} [...] \textit{at some point everything sounded very similar}'' (P4), commented a participant who self-defined herself as a newcomer to EM.
Moreover, another newcomer to EM observed:
``\textit{It was a mentally taxing task because I had to kind of create a small description of every track and compare them in my head}'' (P10).
On the contrary, interviewees affirming to be familiar with EM emphasized how they made use of prior knowledge for categorizing artists and tracks, also discussing how this could have led to some sort of preference or prejudice while evaluating diversity:
``\textit{I can feel like I can make a better decision of what is diverse} [...] \textit{but then there is kind of a bias that comes based on the fact that I like this music a lot}'' (P2).

Here, we observe a contrast between the higher difficulty of the newcomers and the role of prior knowledge for experts as a facilitator to assess diversity. 
On the one hand, newcomers not having prior knowledge could rely more on generic listenable features while interacting with unknown music (e.g. tempo): 
``\textit{People who don't know much about a particular genre, probably would agree on some things that are a bit more generic}'' (P3). 
At the same time, they can associate such features to generic or stereotypical representations of the unfamiliar genre: 
``\textit{[...] this kind of prejudice or bias we have about music that we do not know because we get to know them through these representations of what is considered to be}'' (P12).
On the other hand, experts are assumed to be strongly reliant on their knowledge, which can stimulate inner reflections and thoughts:
``[Experts] \textit{add more layers of abstraction, or more complexities to the assessment of diversity}'' (P2). 
Moreover, they may be more thoughtful to specific differences: ``\textit{If you are familiar with a specific genre, you are more receptive, and you can find distinct patterns}'' (P12).

Interestingly, several interviewees declared how participating in the study made them realize the limits of their knowledge about the diversity of the EM scene, both in terms of musical variation and sociodemographic variation:
``\textit{I realized while making the survey that I had a very strict definition about electronic music myself}'' (P12).
Under this perspective, we observe how the mere exposure to several lists, specifically designed to highlight the diversity of the EM scene, made participants aware of their limited knowledge and stereotypical representation that even unconsciously they may associate with EM. 
This supports the idea that to be exposed to a diverse list of music can be a method to stimulate a self-reflection about prior beliefs, in line with the findings presented by Clarke et al. in \cite{Clarke2015}. 
Indeed, in the survey wherein interviewees participated, we choose to include non-mainstream tracks and artists, especially from groups that are normally underrepresented in the EM scene, pursuing what is defined by Helberger et al. \cite{Helberger2018} as exposure diversity with an adversarial-deliberative perspective. 
Consequently, we provide to the participants an additional viewpoint to reflect on their knowledge: 
``\textit{The electronic music artists that I went to listen to and that I liked before the survey were predominantly white male, which I suppose is still what is predominant in the industry to some extent [...] but definitely it is not the only thing}'' (P13).

Starting from these differences on evaluating the diversity of music lists, we continue discussing the role of diversity with regards to music recommendations and listening practices.

\subsection{Music Recommendation and Diversity}\label{sec:qual2}
Interviewees highlighted how recommendation diversity enables a wider range of choices while listening to music, contrasting the repetitiveness and monotony that eventually could lead to boredom: 
``[Diversity] \textit{gives me more opportunities, not only to discover but also the possibility to choose. If everything sounds similar I may get tired after a bit}'' (P4).
This relationship between diversity, choice difficulty, and overall satisfaction has been already found in several studies in the RS literature \cite{Knijnenburg2012}. 
Nonetheless, there is a contrast between diversity as opposed to monotony in the listening experience, to which participants associate different feelings.

On the one hand, there can be satisfaction in listening to what is considered familiar or expected, because of the immediate reward that can be gained: 
``\textit{I get frustrated when you get caught going in the same loops, but it is at some level also slightly satisfying [...] it is kind of brain-numbing in a way}'' (P2).
A positive aspect linked to the availability of a more diverse set of recommendations is strictly related to the idea of serendipity, i.e. to receive something unexpected but valued positively \cite{Chang2018}: 
``\textit{I ended up once listening to a music that was tagged like Afghan metal, something it was great, I would not imagine searching this type of music}'' (P12).  
However, such a positive vision of the unexpected is not shared by everyone: 
``\textit{I am not the one who likes to take the risks [...], because some genres of music to me are definitely awful, and I do not want to be exposed to that}'' (P11). 
The difficulty of facing something different, or the willingness of taking a risk when leaving the comfort zone for listening to some music somehow considered diverse, are some of what interviewees mentioned as limitations of recommendation diversity: 
``\textit{Taking that plunge into something new is really hard, and I do not know if recommender systems could bridge that barrier of making it more attractive}'' (P13).

Most of the participants also highlighted that in specific moments diversity can be beneficial (e.g. when you want to discover), while in others may be counterproductive (e.g. when you want to focus):
``\textit{If I click on a playlist very homogenous then I would like to continue with that flow. But if it is something more heterogeneous, I can like it the same [...] in the end} [diversity] \textit{could be positive or negative at the same time}'' (P14).
From another perspective, they also affirmed that the most effective way for exposing people to diversity could be by forcing them to take the risk:
``\textit{When you cannot change the playlist or whatever, you know, you are forced to that, and that is the only way that you can listen to something radically new}'' (P11).

Lastly, participants have been asked if they believe that recommendation diversity could be a tool for modifying prior beliefs with regards to a specific musical genre or culture. 
They have also been asked to provide examples when they have experienced such a change in opinion due to the interaction with Music RS. 
As a result, some of the participants presenting their experiences pointed out that, even when they initially disliked or had prejudices, algorithmic recommendations helped them in discovering new facets of previously disliked music genres, which they eventually ended up enjoying and listening to more often:
`\textit{From the idea of classical music that I had [...] quite boring and everything very similar, thanks to some recommendations I have been able to discover different styles and to be in the mood to try to listen to it}'' (P1); 
`\textit{I never liked EDM but [...] algorithms presented to me different tracks, and I found myself listening to it, and noticing the differences within this genre. In the end, I started listening to it more often}'' (P8).
\section{Conclusion and Future Work}\label{sec:conclusions}
The diversity of recommendations can make people aware of the nuances of music previously unknown or unfamiliar, or with a negative attachment, facilitating the discovery and leading to serendipity. 
Nonetheless, as several interviewees mentioned, the benefits resulting from the interaction with algorithmic recommendations can be experienced only if the listener is willing to diversify her listening experience.

In this work, by interviewing listeners we identify different aspects characterizing the diversity assessments, and the interactions with music recommendation diversity. 
The generalizability of such insights is relatively limited by the population of our study, mostly formed by subjects coming from the so-called WEIRD (Western, Educated, Industrialized, Rich, and Democratic) societies \cite{Henrich2010}. 
Replicating the study considering a different sample of listeners both in terms of socio-economic status and origin could strengthen our results, and ideally provide new findings to complement our work.
Even if the presented qualitative analysis highlights some aspects of exposing individuals to diverse music, as future work we foresee exploring more in-depth the dynamics of such interactions designing a longitudinal study, wherein listeners are periodically presented to diversity-aware music recommendations. 
The relevance that simulation-based and longitudinal experiments are gaining in RS research is illustrative of the interest in deepening the long-term impact of such technology \cite{Jannach2020}, and few results have already been presented in the music field (e.g., \cite{Ferraro2020}).

In conclusion, with this work we hope to foster the debate around new practices to assess the impact of music recommendations diversity on listeners, artists, and on our society at large.

\begin{acks}
This work is partially supported by the European Commission under the TROMPA project (H2020 - grant agreement No.\ 770376).
It is also part of the project Musical AI - PID2019-111403GB-I00/AEI/ 10.13039/501100011033 funded by the Spanish Ministerio de Ciencia, Innovación y Universidades (MCIU) and the Agencia Estatal de Investigación (AEI).
This work is also partially supported by the HUMAINT programme (Human Behaviour and Machine Intelligence), Joint Research Centre, European Commission. 
The project leading to these results received funding from "la Caixa" Foundation (ID 100010434), under the agreement LCF/PR/PR16/51110009.
\end{acks}

\bibliographystyle{ACM-Reference-Format}
\bibliography{chiir22-8}

\end{document}